\documentclass{article}
\usepackage{textcomp}
\usepackage{siunitx}
\DeclareSIUnit\permille{\text{\textperthousand}}
\usepackage[utf8]{inputenc}
\usepackage{amsmath}
\makeatletter
\def\tagform@#1{\maketag@@@{[\ignorespaces#1\unskip\@@italiccorr]}}
\makeatother
\usepackage{bm}
\usepackage{authblk}
\usepackage{amssymb}
\usepackage{graphicx}
\DeclareGraphicsExtensions{.png,.pdf,.eps}
\usepackage{ifpdf}
\usepackage{cite}
\usepackage{doi}
\usepackage{tikz}
\usepackage[
	capitalise,
]{cleveref}

\newcommand\norm[1]{\left\lVert#1\right\rVert}

\usepackage{etoolbox}
\usepackage{color}
\renewcommand{\r}{\textcolor{red}}

\graphicspath{{./figures/}}

\newlength\Colsep
\newlength{\savedparindent}

\title{Joint T1 and T2 Mapping\\
	with Tiny Dictionaries\\
	and Subspace-Constrained Reconstruction
}

\author[1]{Volkert Roeloffs}
\author[2,3]{Martin Uecker}
\author[1,3]{Jens Frahm}

\affil[1]{Biomedizinische NMR, MPI für biophysikalische Chemie, 37070 
	Göttingen, Germany}
\affil[2]{Institute for Diagnostic and Interventional Radiology,	University 
	Medical Center, 37075 Göttingen, Germany}
\affil[3]{German Centre for Cardiovascular Research (DZHK), partner site 
	Göttingen, Germany}

\begin{document}
\setlength{\savedparindent}{\parindent}

\maketitle

\vfill

\noindent
{\em Running head:} {T1 and T2 Mapping with Tiny Dictionaries}

\vspace{0.5cm}

\noindent
{\em Correspondence to:} \\
Dr. V. Roeloffs\\
Biomedizinische NMR, MPI für biophysikalische Chemie\\
37070 Göttingen, Germany\\
\texttt{volkert.roeloffs@mpibpc.mpg.de}

\vspace{0.3cm}
\noindent
Approximate word count: 148 (abstract) 2900 (body)\\
Number of pages: 14\\
Number of figures: 5\\
Number of tables: 0\\

\vspace{0.3cm}
\noindent
Submitted to \textit{Magnetic Resonance in Medicine} as a Note\\
\begin{tabular}{@{}lc}
Date of submission: & {2018-12-22} \\
Date of revision: & \r{XXX}
\end{tabular}

\vspace{0.3cm}
\noindent

\newpage
\setlength{\parindent}{0in}
\section*{Abstract}

{\bf Purpose:} 
To develop a method that adaptively generates tiny dictionaries for joint 
$T_1$-$T_2$ mapping.\\
{\bf Theory:}
This work breaks the bond between dictionary size and representation accuracy  
(i) by approximating the Bloch-response manifold by piece-wise linear functions 
and (ii) by adaptively refining the sampling grid depending on the  
locally-linear approximation error.
\\
{\bf Methods:}
Data acquisition was accomplished with use of an 2D radially sampled 
Inversion-Recovery Hybrid-State Free Precession sequence.
Adaptive dictionaries are generated with different error tolerances and
compared to a heuristically designed dictionary. Based on simulation 
results, tiny dictionaries were used for $T_1$-$T_2$ mapping in phantom and in 
vivo studies. Reconstruction and parameter mapping were 
performed entirely in 
subspace.\\
{\bf Results:} 
All experiments demonstrated excellent agreement between the proposed mapping 
technique and template matching using heuristic dictionaries.\\
{\bf Conclusion:}
Adaptive dictionaries in combination with manifold projection allow to reduce 
the necessary dictionary sizes by one to two orders of magnitude.

\vspace{0.5in}
{\bf Key words:} {subspace reconstruction, multi-parametric mapping, T1 
mapping, T2 mapping, model-based, dictionary, quantitative MRI}
\newpage

\setlength{\parindent}{\savedparindent}

\setcounter{secnumdepth}{0}
\section{Introduction}
Multi-parametric mapping of MRI-detectable physical or physiological quantities 
has the potential to detect subtle abnormalities earlier and in a more 
objective manner than conventional contrast-weighted imaging. However, 
traditional methods that acquire a set of fully sampled images with varying 
contrasts and then perform a pixelwise fitting are typically very 
time-consuming. Model-based methods accelerate the measurement by estimating 
the quantitative maps directly from undersampled k-space data and remove the 
need to acquire
fully sampled images \cite{Block2009, Doneva2010, Petzschner2011, Sumpf2011, 
Huang2012,Velikina2013}.
 
Recently, new approaches have been presented that break with 
simple signal models and employ more sophisticated excitation patterns 
\cite{ma2013,Buonincontri2016,Asslaender2017,Asslaender2017a,
Asslaender2018,Zhao2016}. One way to deal with the resulting complex signal 
responses is to generate a 
bank of signal prototypes or “dictionaries”  
\cite{Doneva2010,ma2013}. However, these dictionaries (i) are typically very large in size, (ii) scale exponentially with the number of parameters, (iii) take long to compute, and (iv) result in a huge number of comparisons at the stage of matching. A variety 
of ideas have been presented to overcome the associated difficulties, as a 
simple reduction of the sampling density would result in a reduction of 
representation accuracy. For example, dictionaries compressed by singular value 
decomposition (SVD) exploit redundancies to perform the matching process in a 
reduced-dimensional space \cite{McGivney2014,yang2018}. With the use of 
clustering properties \cite{Cauley2015} matching can further be sped up as unnecessary comparisons are avoided. Both approaches rely on dictionaries, in which, first of all, sampling positions have been chosen heuristically.

Here, we present a new approach to automatically generate sampling positions in 
an adaptive way. These positions are then considered a set of support 
points that approximate the Bloch-response manifold \cite{Davies2014} by 
piece-wise linear functions. Manifold projection in combination with these 
adaptively designed dictionaries allows reduction of necessary dictionary sizes 
by one to two orders of magnitude. The new method is applied to accomplish 
joint $T_1$ and $T_2$ mapping.

\section{Theory}

The MRI signal response to a complex excitation pattern is given by the Bloch 
equations \cite{Bloch1946}. If the excitation sequence is sufficiently rich and 
the signal response sensitive to the parameter of interest, all signal responses lie on a 
non-linear smooth manifold that is embedded within the higher-dimensional (time-domain) space. The low-dimensional manifold is called the Bloch-response 
manifold \cite{Davies2014} and here used in two ways to break the bond between dictionary size and representation accuracy: First, we approximate the Bloch-response 
manifold by piece-wise linear functions and consider the dictionary a set of 
support points. As a consequence, mapping to the parametric domain becomes 
continuous rather than discretized by the chosen sampling grid. Second, we 
allow the sampling grid to be refined adaptively during the generation of the 
dictionary depending on the precision needed. To this end, an initial grid is 
recursively refined in regions where the locally-linear approximation is not 
accurate enough.

\subsection{Piece-wise linear approximation and adaptive sampling}
The basic idea of an adaptively refined dictionary generation is to allow 
coarse sampling in regions with locally-linear signal dependency and fine 
sampling where non-linear dependencies are present.
More specifically, in the vicinity of a reference position 
$x^\text{ref}=(T_1,T_2)^{\top}$ in $T_1$-$T_2$ parameter space, the 
locally-linear approximation
$ y - y^\text{ref} \approx J(x-x^\text{ref})$ holds, where 
$J$ is the Jacobian matrix (defining the best linear approximation of the 
nonlinear map at position $x^\text{ref}$), $y^\text{ref}$
the signal response at the reference position, and $x$ and $y$ neighboring 
vectors in parameter domain and temporal domain, respectively.
Considering neighborhoods $X$ and $Y$, i.e. matrices with columns being 
neighboring vectors in parameter domain and temporal domain, respectively,  
approximation errors 
\begin{equation}
E_{t n} = \sum _j J_{t j}(X_{j n} - x^\text{ref}_j) - (Y_{t n}-y^\text{ref}_t)
\end{equation}
can be defined for each time point $t$ for each vector $n$ in 
these neighborhoods. The closer the neighboring vectors are to the reference 
position, the smaller the approximation errors become (smoothness of the 
Bloch-response manifold).  

This motivates our proposed strategy for adaptive 
dictionary generation: Starting with an initial neighborhood, the entire 
dictionary can be built by recursively splitting into downsized neighborhoods 
until the total approximation error
$E^\text{tot}=\sqrt{\sum_{t,n} E_{t n}^2}$ fulfills the stopping criterion 
$E^\text{tot}~<~\varepsilon~\sqrt{\sum_{t,n} Y_{t n}^2}$ 
which is controlled by the predefined error tolerance $\varepsilon$.
 
While downsized neighborhoods could be generated in different ways (quadtree- 
or binary space partitioning, isotropic downscaling of neighborhoods, etc.), we 
propose to split the current neighborhood only into a single direction at a 
time. 

More specifically, we identify the index $n^*$ of the neighbor exhibiting the 
largest 
root-sum-squares error
\begin{equation}
n^* = \arg\max_{n} \sqrt{\sum_t E^2_{t n}} 
\end{equation}
and select a coordinate axes as split direction by computing the largest 
relative parameter deviation with respect to the reference position
\begin{equation}
j^* = \arg\max_{j} \left|{(X_{j n^*} - 
x^\text{ref}_j)/x^\text{ref}_j}\right|\quad.
\end{equation}
The current neighborhood is then split into the direction of the $j^*$-th 
coordinate axis yielding two half-sized neighborhoods both subject to recursive 
splitting.

The final result of the recursive building process is a lookup table linking 
$N$ model signals in the dictionary $D_i \in \{1,2,...,N\}$  to their 
corresponding position in parameter space. Here, for the proposed manifold 
projection, the Jacobian matrix is stored additionally for each position.

\subsection{Generating neighborhoods}
For joint $T_1$-$T_2$ mapping, the embedded Bloch-response manifold is 
two-dimensional. Consequently, a minimum of two neighbors have to be generated 
for a new reference position in parameter domain. Here, these two neighbors are 
generated according to 
\begin{equation} 
\begin{aligned}
X_{\bullet 1}&=x^\text{ref} 
+ \begin{pmatrix}   2^{-p}\Delta T_1 \\ 2^{-q} \Delta T_2   \end{pmatrix}\\
\text{ and }X_{\bullet 2}&=x^\text{ref} 
+ \begin{pmatrix}   2^{-p}\Delta T_1 \\ 0   \end{pmatrix},\quad
\text{where } \Delta T_{1/2}=T^\text{max}_{1/2}-T^\text{min}_{1/2}.
\end{aligned}
\end{equation}
The integers $p$ and $q$ reflect the recursion depths in $T_1$ and $T_2$ 
splitting direction, 
respectively, and are increased as long as the approximation error exceeds the 
prescribed threshold.

\subsection{Manifold projection}

As the final approximation error for all neighborhoods is smaller than the 
error tolerance, it is guarantied that the non-linear Bloch-response in the 
vicinity of each entry $D_i$ can linearly be approximated by the respective 
Jacobian matrix. 
Consequently, the manifold projection is realized in two steps. First, an 
appropriate entry $y^{\text{ref}}$ in the dictionary has to be identified which 
is realized by pattern matching similar to Refs. 
\cite{ma2013,Jiang2015,yang2018}: 
\begin{equation}
\begin{aligned}
y^{\text{ref}}&=D_{i^*}, \text{ where } 
i^*=\arg\max_i\frac{\left|<D_i,y>\right|}{{\norm{D_i}}_2}
\end{aligned}
\end{equation}
Second, the linear function in the region around this reference position has to 
be inverted to project the reconstructed response signal $y$ onto the 
piece-wise linear manifold in the parametric domain. The final projection can 
be cast into a least-squares 
optimization problem of the form
\begin{equation}
\label{eq:projection}
\begin{aligned}
\{\hat{x},\hat{\rho}\}&=\arg\min_{x,\rho} 
{\norm{J(x-x^{\text{ref}})+y^{\text{ref}}-\rho^{-1} y}}_2^2\\
&=\arg\min_{x,\rho} 
\norm{{A(x,\rho^{-1})^\top-Jx^{\text{ref}}+y^{\text{ref}}}}_2^2,
\quad \text{where } A:=[J|-y]
\end{aligned}
\end{equation}
and can be solved by the Moore-Penrose pseudo inverse $A^+$ to yield the final 
 quantitative result 
$(\hat{T_1},\hat{T_2},\hat{\rho}^{-1})^\top=A^+(Jx^{\text{ref}}-y^{\text{ref}})$
which includes the proton density $\rho$. This  
scaling constant is added as an additional unknown as all entries in the 
dictionary have been generated with unit proton density.

\section{Methods}

\subsection{Heuristic design and template matching}

To evaluate accuracy and size of the adaptively generated dictionaries, the 
heuristically designed dictionary in Ref. \cite{ma2013} is chosen as a 
benchmark. Ma and coworkers partitioned the $T_1$-$T_2$ space into 4 regions 
with 
different sampling densities (see \Cref{Figure 1}{E}). The parametric 
representation of a signal response is found by identifying the best matching 
entry in the dictionary $D$ (template matching) and assigning the corresponding 
parameter values from the lookup table to this pixel. This procedure is 
identical to the first step of 
our proposed manifold projection. 

For a meaningful comparison, all adaptive dictionaries share the boundary 
conditions of Ref. \cite{ma2013}, namely
$T_1 \in [\SI{0.1}{\s},\SI{5}{\s}]$,
$T_2 \in [\SI{0.02}{\s},\SI{3}{\s}]$,
and the physical constraint $T_1 \ge T_2$.
Here, offsets in the $B_0$ field are excluded.

\subsection{MRI}
All MRI studies were performed at a field strength of \SI{3}{T} (Magnetom 
Prisma, Siemens Healthineers, Erlangen, Germany) using a 64-channel head coil. 
Volunteers without known illness were recruited and written informed consent 
was obtained before MRI according to the regulations of the local ethics 
committee. 

Data acquisition was accomplished with use of an Inversion-Recovery (IR) 
Hybrid-State Free Precession (HSFP) experiment \cite{Asslaender2018} to 
sensitize the MRI response signal to $T_1$ and $T_2$ relaxation. The flip angle 
pattern is originally optimized for maximal mapping efficiency at
$T_1=\SI{781}{\ms},T_2=\SI{65}{\ms}$ with 
$T_R=\SI{4.5}{\ms}$. Due to specific absorption rate constraints of the slice-selective excitation pulses, we prolonged the repetition time in this study to $T_R=\SI{5}{\ms}$ 
which scales the expected maximal efficiency to be achieved at 
$T_1=\SI{868}{\ms},T_2=\SI{72}{\ms}$. The flip angle pattern was implemented 
 with 2D radial sampling using a tiny-Golden-Angle scheme \cite{Wundrak2015} 
 with $\Psi_{N=10}\approx \ang{16.9523}$. 
Phantom and brain imaging was performed with a spatial resolution of 
$\SI{0.75x0.75x4}{\mm}$ in a total acquisition time of
$T_{\text{ACQ}}=\SI{4.3}{\s}$. Signal time courses and corresponding gradients  
were computed using the analytical expression for HSFP (compare eq. 7 in Ref 
\cite{Asslaender2018}). Slice profile effects were explicitly taken into 
account  
by evaluating this expression with different $B_1$ strengths. To this end, the 
time course of the implemented RF excitation pulse (bandwidth-time-product  
2) was Fourier transformed and one of the two symmetric lobes discretized into 
20 factors that scale each flip angle in the excitation pattern. The 
slice-profile compensated signal time course was eventually obtained by 
averaging the 20 individual time courses.

\subsection{Subspace and reconstruction}
To further reduce dictionary size and to minimize noise amplification, we 
formulate the reconstruction as a subspace-constrained \cite{Petzschner2011, 
Huang2012} linear inverse problem. 
A subspace size of $K=4$ was chosen heuristically and the subspace basis was 
determined by performing a SVD \cite{Petzschner2011, McGivney2014, Tamir2016} 
on either the full adaptive dictionary (phantom and brain study) or on a set of signals from a uniform grid in $T_1$-$T_2$ parameter space (numerical simulation). The latter strategy ensures a dictionary-independent basis. With this choice, the following 
minimization problem is solved:
\begin{equation}
\alpha^* = \arg\min_{\alpha} {\norm{y-\mathcal{P}_{\vec{k}}\mathcal{F}S\Phi_K 
\alpha}}_2^2 + 
\lambda R(\alpha)
\label{eq_lin_inv_prob}
\end{equation}
where $y$ denotes the radial raw data, $\mathcal{P}_{\vec{k}}$ the projection 
onto the sampled k-space trajectory, $\mathcal{F}$  the Fourier transform, $S$ 
multiplication with the (predetermined) coil sensitivity profiles, $\Phi_K$ 
the temporal basis, and $\alpha$ the unknown subspace coefficients. Coil 
sensitivity profiles $S$ were predetermined by ESPIRIT \cite{Uecker2014} using 
the gridding solution of the first subspace coefficient, and spatial 
correlations across subspace coefficients were exploited by a locally-low-rank 
regularizer $R$ \cite{Tamir2016, Roeloffs2018}.

Similar to our previous work \cite{Roeloffs2018}, gridding, gradient delay 
correction, and precomputation of the transfer point-spread-function is 
performed by custom MatLab routines, while image reconstruction was performed 
by a customized version of BART \cite{BART} using the ADMM optimizer 
\cite{Boyd_2010} ($\rho=0.01$, 100 iterations) and locally-low-rank 
regularization ($\lambda=0.0003$ and block size $8\times8$).

The linear subspace transformation $\Phi^\top_K$ is also applied to each entry 
in the dictionary $D$ and each Jacobian matrix $J$, such that the manifold 
projection 
is simply performed with their subspace representations $\hat{D}=\Phi^\top_K D$ 
and $\hat{J}=\Phi^\top_K J$.

\subsection{Phantom design}
For a quantitative validation, a home-brew phantom was designed consisting of 9 
gel tubes with distinct $T_1$ and $T_2$ values. Closely following Ref. 
\cite{Hattori2013}, $\mathrm{GdCl_3}$ was used as a $T_1$ 
modifier and agarose as a $T_2$ modifier to generate $T_1$ and $T_2$ values in 
the range of typical relaxation times for white and gray matter.  
To access the power of separability of the proposed method, $T_1$ was kept 
approximately constant while varying $T_2$ and vice versa. Ground truth $T_1$ 
relaxometry was realized by four IR single-echo spin-echo data acquisitions 
\cite{Stikov2015} (TI=\SIlist{30;530;1030;1530}{\ms}) and pixel-wise fitting 
of the complex data using a freely available custom software package 
\cite{Barral2010}. $T_2$ gold standard values were obtained by 5 single-echo 
spin-echo data acquisitions (TE = \SIlist{12;30;73;182;450}{\ms}, 
TR=\SI{4.5}{\s}) and subsequent fitting of a mono-exponential model to the 
magnitude data.

\subsection{$B_1$ profile correction}
Local deviations in the $B_1$ field are known to be a major source of systematic 
errors in quantitative MRI. This specifically applies to the utilized HSFP 
sequence as information about $T_1$ and $T_2$ is encoded in the signal response 
by traversing the Bloch sphere on a particular path \cite{Asslaender2018}. 
Imperfect $B_1$ field strength leads to deviations from the intended path and 
mainly results in inaccurate $T_2$ values, similar to $B_1$ effects in "MR
Fingerprinting" \cite{Buonincontri2016,Ma2017}. 
To correct for this $B_1$ deviations, a separate $B_1$ map was acquired for the 
phantom study using a standard sequence of the vendor (Bloch-Siegert method 
\cite{sacolick2010}) which matched the spatial resolution of the HSFP sequence.
The pixel-wise information about the relative scaling of the nominal $B_1$ 
strength, $rB_1$, was used to correct at the stage of the manifold projection. 
In the spirit of the local-linear approximation, the computed Jacobian matrices 
can be 
extended to incorporate the derivative with respect to $rB_1$ and 
\Cref{eq:projection} is extended to provide signal models for different $B_1$ 
strength. 

\subsection{Code availability}
The source code will be made publicly 
available via \url{https://github.com/} at the time of publication. 

\section{Results}
\subsection{Simulation}
\begin{figure}[p]
	\centering
	\includegraphics[width=\textwidth]{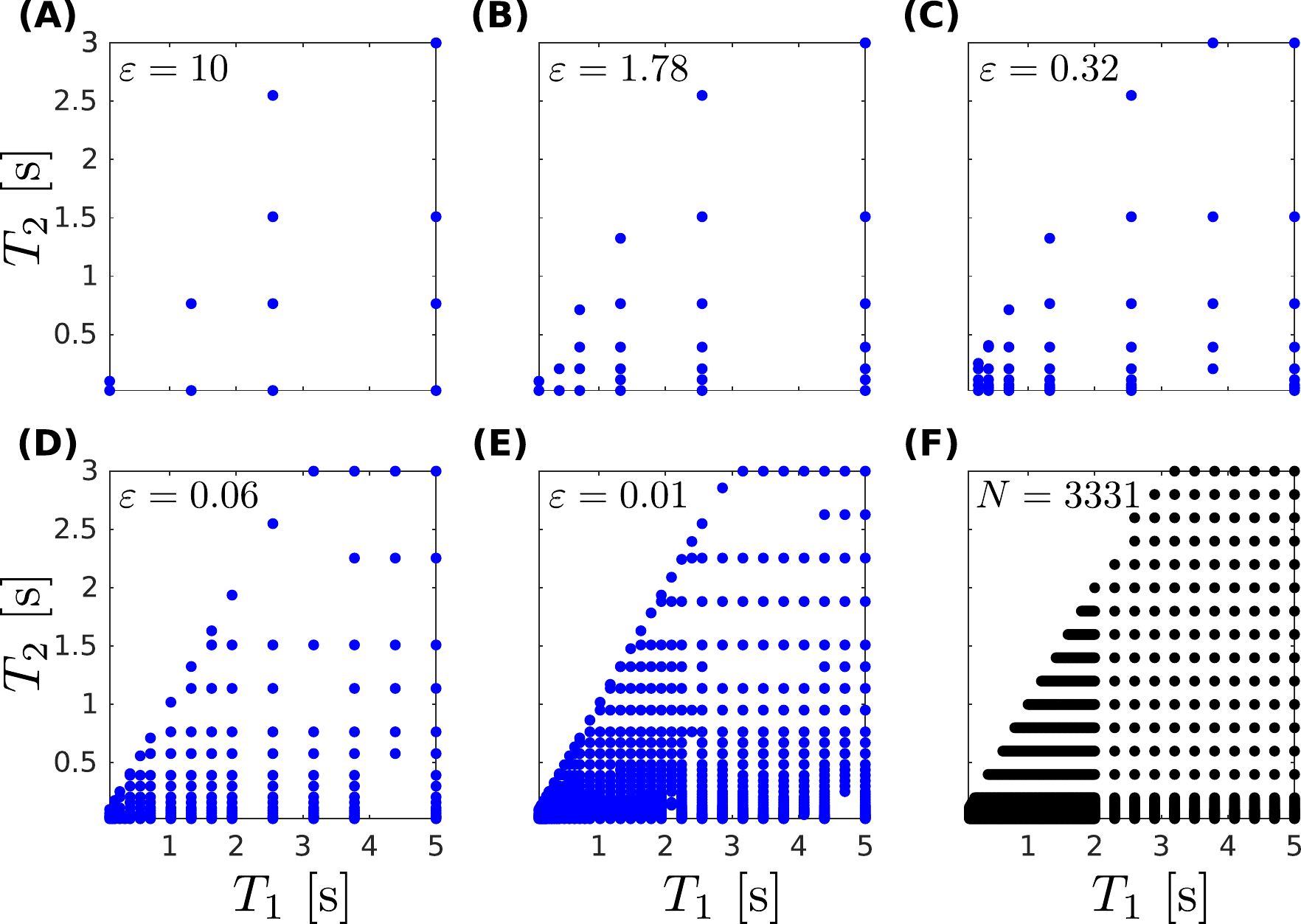}
	\caption{\label{Figure 1}
		(A-E) Sampling positions in parameter space of dictionaries generated adaptively as a function of error tolerance $\varepsilon$. (F) Sampling positions of the heuristically designed dictionary.
		}
\end{figure}

To investigate the role of the error tolerance $\varepsilon$, adaptive 
dictionaries were generated for decreasing values of $\varepsilon$ and compared 
to the heuristic dictionary. With decreasing $\varepsilon$, the number of 
dictionary entries increases (\Cref{Figure 1}A-E) and the intended adaptivity 
effect becomes apparent: In each dictionary, the sampling density increases 
toward the short-$T_1$-short-$T_2$ region. Note the close similarity between 
the automatically and the heuristically generated density distributions 
(\Cref{Figure 1}F).

\begin{figure}[p]
	\centering
	\includegraphics[width=0.5\textwidth]{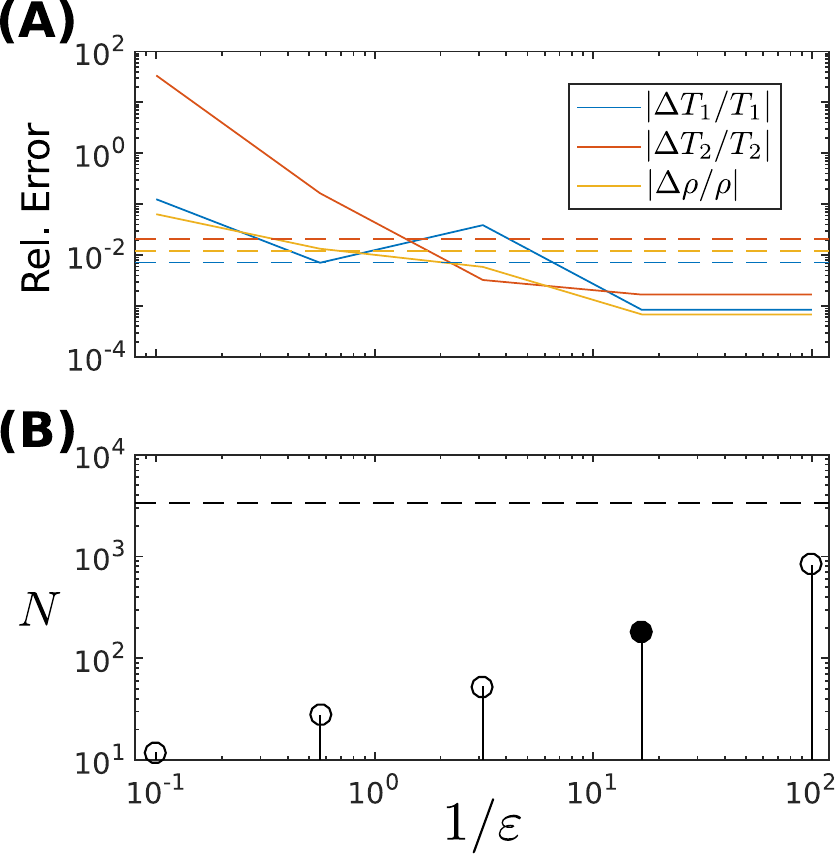}
	\caption{\label{Figure 2}
		(A) Relative error in $T_1$, $T_2$, and $\rho$ as a function of the inverse error tolerance when projecting a probing signal using the adaptive dictionaries in \Cref{Figure 1} (solid lines). For comparison, the result of template matching with the heuristic dictionary (dashed lines) is shown.
		(B) Number of entries $N$ in the adaptive dictionaries as a function of 
		the inverse error tolerance. Size of the heuristic dictionary (dashed 
		line) for reference. The dictionary generated with $\varepsilon=0.06$ 
		and $N=181$ (solid circle) was used for phantom and in vivo studies. 
		}
\end{figure}

In \Cref{Figure 2}, these dictionaries have been used to project a probing 
signal response ($T_1=\SI{1.088}{\s}$, $T_2=\SI{0.069}{\s}$, $\rho=\num{1}$) 
that is not contained in any of the dictionaries. With decreasing 
$\varepsilon$, the relative error in $T_1$, $T_2$, and $\rho$ generally 
decreases and finally falls below the heuristic error and levels below 
\SI{1.1}{\permille}. Comparing the  dictionary sizes (\Cref{Figure 2}B) reveals 
that the adaptive sampling strategy, depending on the chosen error tolerance, 
results in dictionaries reduced in size by one to two orders of magnitude 
compared to heuristic sampling. Based on the excellent accuracy obtained with 
only 181 dictionary entries, the error tolerance of $\varepsilon=\num{0.06}$ 
was used for both phantom and in vivo studies. 

\subsection{Phantom experiments}

\begin{figure}[p]
	\centering
	\includegraphics[width=\textwidth]{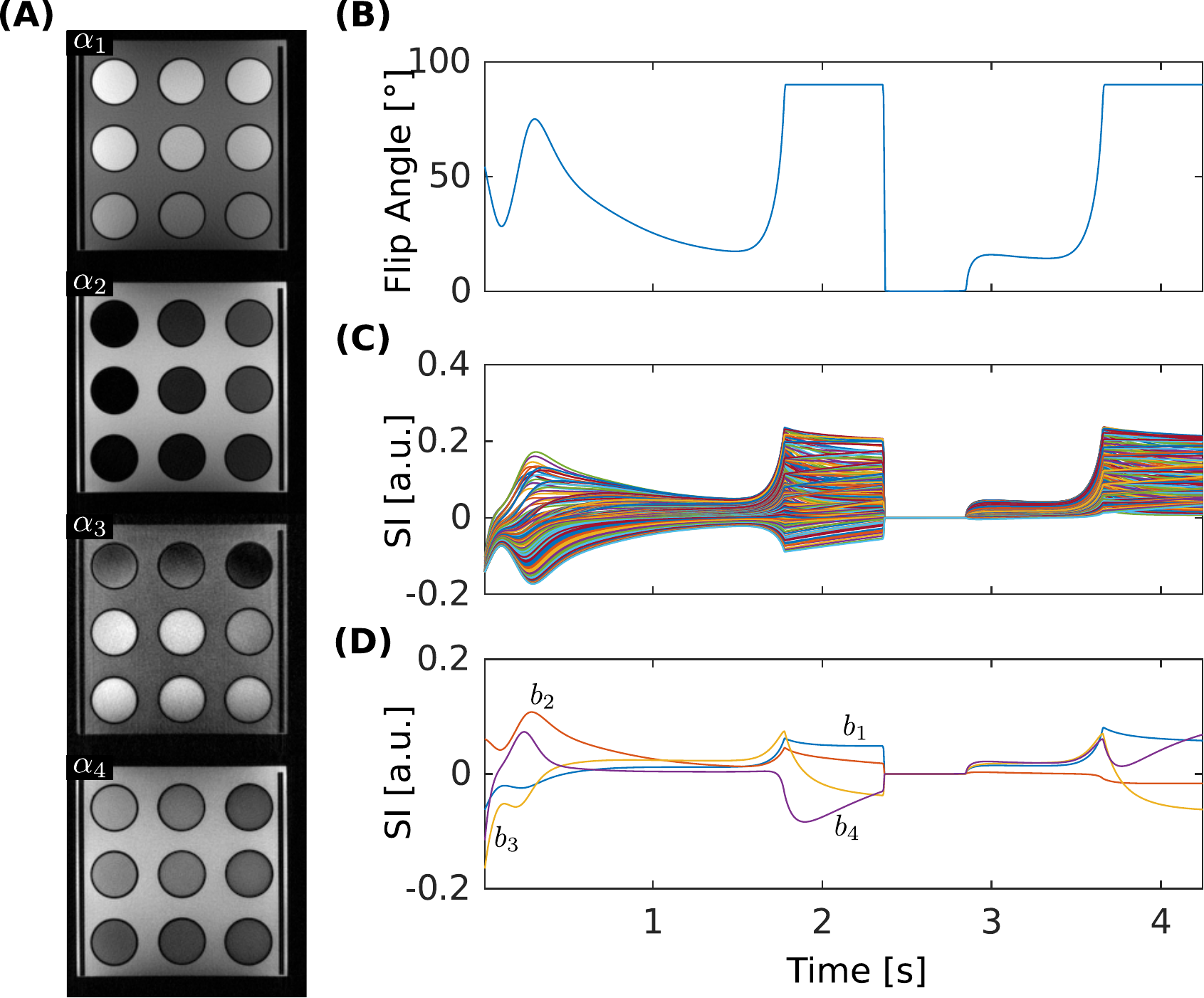}
	\caption{\label{Figure 3}
		(A) Reconstructed subspace coefficient maps, (B) implemented HSFP excitation pattern, (C) visualization of all signal time courses contained in the dictionary, and (D) basis functions as obtained after SVD of the full dictionary. Coefficient $\alpha_i$ refers to basis function $b_i$.
		}
\end{figure}

\Cref{Figure 3}A shows the four reconstructed subspace coefficient maps  
obtained for the $T_1$-$T_2$ phantom using the HSFP flip angle pattern 
(\Cref{Figure 3}B). The subspace approach allows to store the signal responses for all 
sampled $T_1$-$T_2$ parameter combinations (\Cref{Figure 3}C) in a 
compressed representation with four coefficients per sampling point in the 
dictionary (basis functions shown in \Cref{Figure 3}D).

\begin{figure}[p]
	\centering
	\includegraphics[width=\textwidth]{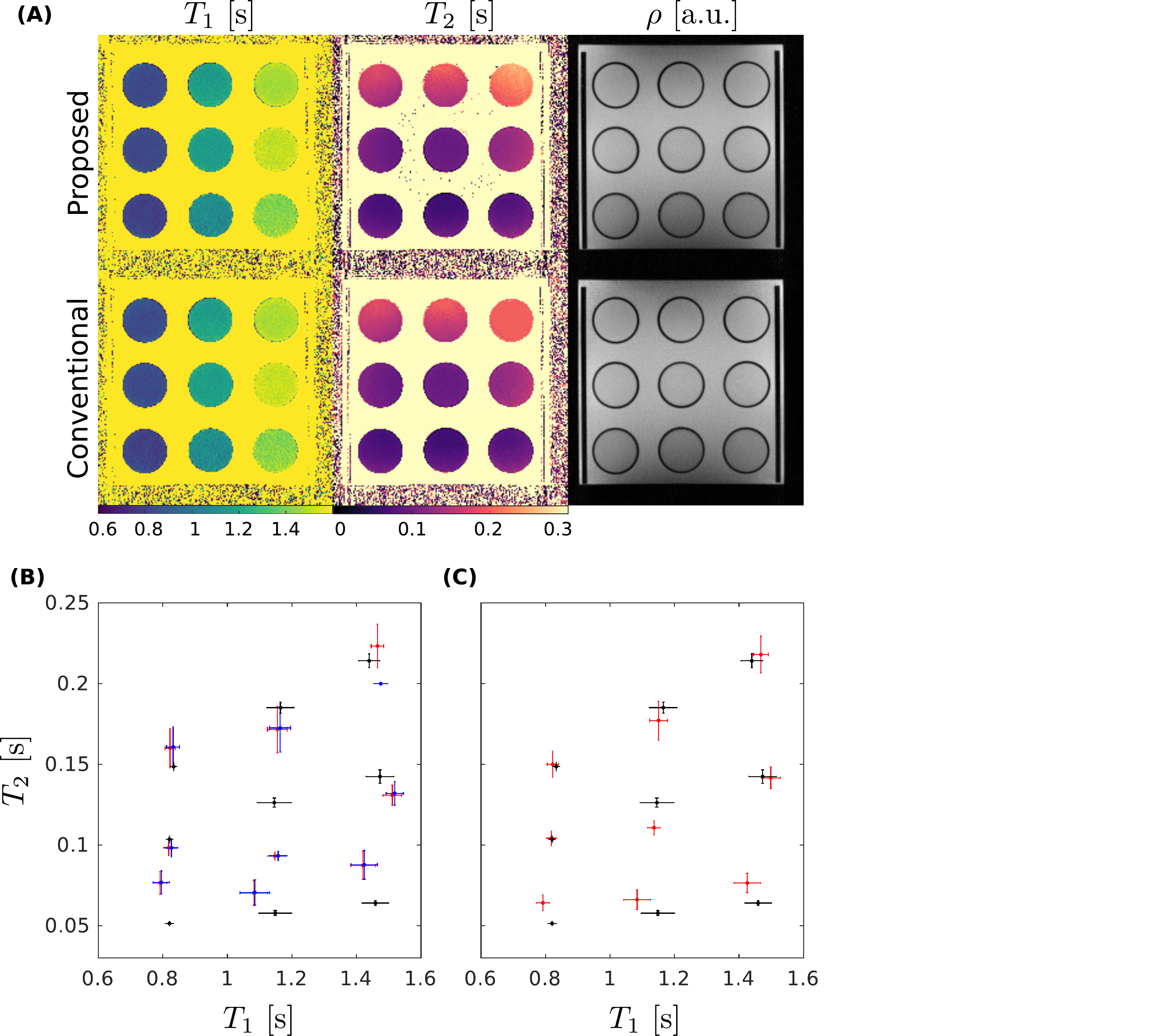}
	\caption{\label{Figure 4}
		Parameter maps and corresponding ROI mean and standard deviation 
		obtained by the proposed method (upper row in (A) and red crosses in 
		(B-C)) in comparison to template matching with heuristic dictionary 
		(lower row in (A) and blue crosses in (B)). (C) Proposed method with 
		$B_1$ profile correction. Black crosses in (B, C) represent values 
		obtained by IR/multi-echo spin-echo MRI.
		}
\end{figure}

The reconstructed subspace coefficients are then mapped pixel-wise to yield the 
final $T_1$, $T_2$ and proton density maps (\Cref{Figure 4}). This is done by 
the proposed manifold projection using the adaptively generated dictionary 
(\Cref{Figure 4}A), as well as by template matching with the heuristic 
dictionary (\Cref{Figure 4}B). Quantitative comparison of ROI-wise mean and 
standard deviations shows excellent agreement between these two methods except 
for the longest-$T_1$-longest-$T_2$ tube (upper right grid position). Here, the 
template matching approach leads to a "cartoon" artifact in the $T_2$ map. The 
entire compartment is mapped to a constant value of \SI{0.2}{\s} 
resulting in a vanishing standard deviation. Sampling was obviously too coarse 
in the heuristic dictionary in the region around $T_2=\SI{0.2}{s}$ (see 
\Cref{Figure 1}E).

The quantitative values are in general agreement with the gold standard 
measurements, however, for both mapping methods a $T_2$ bias is noticeable. 
Applying the proposed $B_1$ profile correction in the manifold projection 
(\Cref{Figure 4}C) corrects this bias to a large degree.

\subsection{In Vivo experiments}
\begin{figure}[p]
	\centering
	\includegraphics[width=\textwidth]{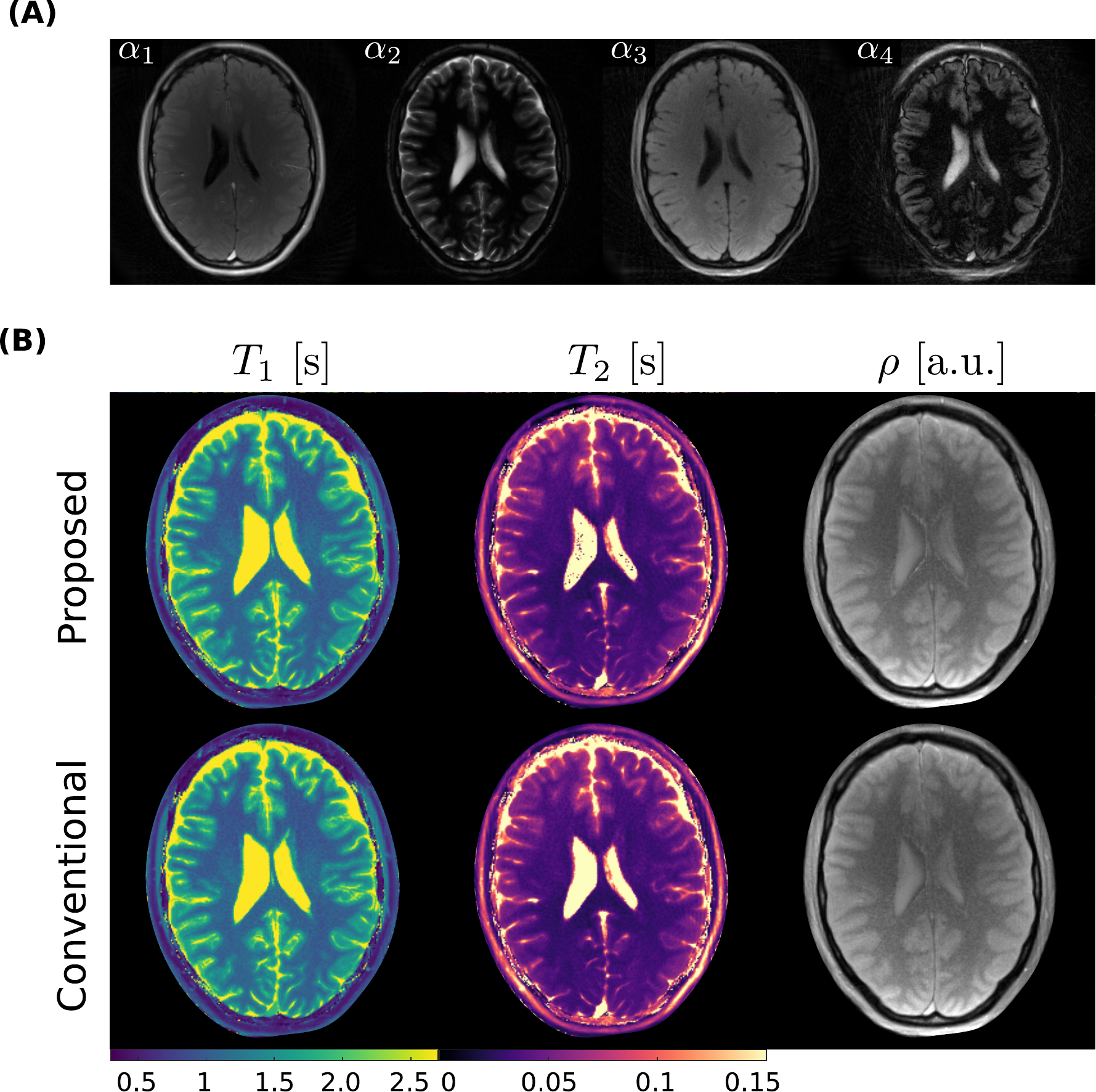}
	\caption{\label{Figure 5}
		(A) Subspace coefficient maps and (B) corresponding parameter maps for a transverse section of the human brain in analogy to \Cref{Figure 4}. The parameter maps are without $B_1$ correction and masked to the region of image support.
		}
\end{figure}

\Cref{Figure 5} shows subspace coefficients and parameter maps of a transversal section of the human brain. The parameter maps reveal excellent agreement and demonstrate the efficient use of tiny dictionaries for multi-parametric mapping in vivo.

\section{DISCUSSION}

In this work, the locally-linear model for joint $T_1$ and $T_2$ mapping was built with the analytical Jacobian. Depending on the employed signal model (extended phase 
graphs (EPG), full Bloch simulation, etc.), this computation can be cumbersome. 
In this case, a proper replacement for the exact Jacobian matrix is required in 
the 
proposed manifold projection. After identification of a specific neighborhood, 
an approximate Jacobian matrix can easily be obtained by linear regression 
analysis using the reference position and its neighborhood. With this 
approximation, the proposed manifold mapping is also applicable to cases in 
which analytical information on signal 
derivatives is not easily available.

The quantitative evaluation of the $T_1$-$T_2$ phantom demonstrated that the 
proposed $B_1$ correction removes the original $T_2$ bias to a large degree. 
However, in particular the short-$T_2$ compartments showed a remaining bias 
which can probably be attributed to neglected effects in the signal 
model such as finite RF pulses, $T_2$-dependent inversion efficiency, and magnetization transfer.

Here, we implemented the HSFP excitation pattern in a 2D (rather than 3D) sequence, so that the signal response becomes a through-slice average. While 
a proper compensation in the forward model was possible, the excitation pattern 
is suboptimal in terms of mapping efficiency. Although a rigorous optimization 
including the slice profile was beyond the scope of this work, it would 
increase the mapping efficiency.

In contrast to non-linear model-based reconstruction techniques, linear 
subspace-constrained techniques are inherently tolerant to partial 
voluming and allow fast reconstruction. However, the choice of the subspace 
size always becomes a trade-off between noise amplification and model-error. 
Therefore, it would be highly desirable to combine the following techniques: A 
non-linear signal model for optimal use of data redundancy, and its embedding 
in a linear subspace for computational efficiency would eventually 
make the noise amplification independent of the subspace size. The proposed 
manifold projection could constitute a key role in such a fused reconstruction 
technique, but further investigations are necessary. 

In conclusion, a novel method to adaptively generate dictionaries 
for multi-parametric mapping was introduced. The quantitative results for 
$T_1$-$T_2$ mapping showed excellent agreement between the proposed manifold projection using adaptive dictionaries and template matching using heuristic dictionaries. The demonstrated ability to perform reconstruction and parameter mapping entirely in subspace justifies the coined term "tiny dictionaries". The proposed 
technique has the potential to overcome problems associated with large 
dictionaries in quantitative multi-parametric mapping.

\clearpage
\bibliographystyle{mrm}
\bibliography{biblio}

\end{document}